\documentclass[conference]{IEEEtran}
\IEEEoverridecommandlockouts

\usepackage{cite}
\usepackage{amsmath,amssymb,amsfonts}
\usepackage{algorithmic}
\usepackage{graphicx}
\usepackage{subfig}
\usepackage{textcomp}
\usepackage{xcolor}
\usepackage{multicol}
\usepackage{multirow}
\usepackage{threeparttable}
\usepackage{hyperref}

\graphicspath{{figures/}}
\begin{document}

    \title{EDCNN: Edge enhancement-based Densely Connected Network with Compound Loss for Low-Dose CT Denoising}

    \author{
        \IEEEauthorblockN{Tengfei Liang}
        \IEEEauthorblockA{\textit{Institute of Information Science} \\
            \textit{Beijing Jiaotong University} \\
            Beijing, China \\
            tengfei.liang@bjtu.edu.cn
        } \\
        \IEEEauthorblockN{Tao Wang}
        \IEEEauthorblockA{\textit{Institute of Information Science} \\
            \textit{Beijing Jiaotong University} \\
            Beijing, China \\
            twang@bjtu.edu.cn
        } \and
        \IEEEauthorblockN{Yi Jin}
        \IEEEauthorblockA{\textit{Institute of Information Science} \\
            \textit{Beijing Jiaotong University} \\
            Beijing, China \\
            yjin@bjtu.edu.cn
        } \\
        \IEEEauthorblockN{Songhe Feng}
        \IEEEauthorblockA{\textit{Institute of Information Science} \\
            \textit{Beijing Jiaotong University} \\
            Beijing, China \\
            shfeng@bjtu.edu.cn
        } \and
        \IEEEauthorblockN{Yidong Li}
        \IEEEauthorblockA{\textit{Institute of Information Science} \\
            \textit{Beijing Jiaotong University} \\
            Beijing, China \\
            ydli@bjtu.edu.cn
        } \\
        \IEEEauthorblockN{Congyan Lang}
        \IEEEauthorblockA{\textit{Institute of Information Science} \\
            \textit{Beijing Jiaotong University} \\
            Beijing, China \\
            cylang@bjtu.edu.cn
        }
    }

    \maketitle
    \bstctlcite{IEEEexample:BSTcontrol}

    \begin{abstract}
        In the past few decades, to reduce the risk of X-ray in computed tomography (CT), low-dose CT image denoising has attracted extensive attention from researchers, which has become an important research issue in the field of medical images. 
        In recent years, with the rapid development of deep learning technology, many algorithms have emerged to apply convolutional neural networks to this task, achieving promising results.
        However, there are still some problems such as low denoising efficiency, over-smoothed result, etc. 
        In this paper, we propose the \emph{Edge enhancement based Densely connected Convolutional Neural Network (EDCNN)}. 
        In our network, we design an edge enhancement module using the proposed novel trainable Sobel convolution. Based on this module, we construct a model with dense connections to fuse the extracted edge information and realize end-to-end image denoising.
        Besides, when training the model, we introduce a compound loss that combines MSE loss and multi-scales perceptual loss to solve the over-smoothed problem and attain a marked improvement in image quality after denoising.
        Compared with the existing low-dose CT image denoising algorithms, our proposed model has a better performance in preserving details and suppressing noise.
    \end{abstract}

    \bigskip


    \section{Introduction}
        Computer tomography (CT)\cite{buzug_computed_2011} plays a very important role in modern medical diagnosis.
        Regarding its imaging principle, it uses the X-ray beam to scan a certain part of the human body. 
        According to the different absorption and transmission rate of X-ray in different tissues of the human body, it detects and receives the signals passing through the human body with highly sensitive instruments. 
        After conversion and computer processing, the tomographic image of the body to be examined can be obtained.
        Due to the X-ray used in this technology, the potential safety hazard in the radiation process has also caused more and more people's attention and concern\cite{brody_radiation_2007,donya_radiation_2014,10.1001/archinternmed.2009.427,hobbs_physician_2018}.
        
        When performing the CT scan, it will involve the intensity (or dose) of the used X-ray\cite{PMID:18046031}. 
        As demonstrated in \cite{naidich_low-dose_1990}, researchers have found that the higher the dose of X-ray within a certain range, the higher the image quality of the CT image. 
        However, patients will get more potential harm to their bodies with a greater intensity of X-ray. 
        On the contrary, using the lower dose of radiation can reduce safety risks, but it will introduce more image noise, which brings more challenges to the doctor’s later diagnosis.
        In this context, low-dose CT (LDCT) image denoising algorithms are proposed to solve this contradiction. 
        The main idea\cite{kang_deep_2017}\cite{Chen_2017} is that they firstly use CT images under the low-dose radiation as the input of the designed algorithm, and then the algorithm will output noise-reduced CT images. 
        In this way, both radiation safety and CT image quality can be considered at the same time.

        In recent years, through researchers' experiments\cite{kang_deep_2017}\cite{Chen:17}, convolutional neural network (CNN) has been shown to have good potential to solve the image denoising task and can achieve better performance than traditional methods.
        As for existing CNN image denoiser, researchers in this field have designed a variety of different structures of models, including fully connected convolutional neural networks (FCN)\cite{Chen:17,Yang_2018,choi_real-time_2018}, convolutional encoder–decoder networks with residual connections\cite{Chen_2017}\cite{hu_artifact_2019} or conveying-paths\cite{Shan_2018,yi_sharpness-aware_2018,gholizadeh-ansari_deep_2020} and some network variants using 3D information\cite{Shan_2018}\cite{7934380}, etc.
        
        Although there have been many models and algorithms, the task of low-dose CT image denoising has not been completely solved. 
        Existing models also face some problems such as over-smoothed results, loss of the edge, and detail information.
        Therefore, how to improve the low-dose CT image quality after denoising is still a key issue that needs to be resolved by researchers.
        In order to have better preservation of image subtle structures and details after the process of noise reduction, our paper proposes a novel CNN model, the \textbf{\emph{Edge enhancement based Densely connected Convolutional Neural Network (EDCNN)}}. 
        The EDCNN is designed as an FCN structure, which can effectively realize the low-dose CT image denoising in the way of post-processing.
        And experiments show that we can get better output results by using this proposed denoiser. 
        In general, the contributions of this paper are summarized as follows:
        
        \begin{itemize}
            \item Design an edge enhancement module based on the proposed trainable Sobel convolution, which can extract edge features adaptively during the optimization process.
            \item Construct a fully convolutional neural network (EDCNN), using conveying-paths densely connection to fuse the information of input and edge features.
            \item Introduce the compound loss used for the training stage, which integrates the MSE loss and multi-scales perceptual loss to overcome over smoothing problems.
        \end{itemize}

        This paper's structure is organized as follows: 
        Section II mainly surveys the related research, including the existing models' composition and structure as well as the mainstream loss function. 
        Section III introduces the designed EDCNN model and explains the contribution of this paper in terms of method. 
        In section IV, we show the experimental configuration and the corresponding experimental results. 
        In the end, section V makes a comprehensive summary of our work.

    \section{Related Work}
        In this section, we show the existing methods related to the low-dose CT image denoising task and discuss their implementation and performance.

        \medskip

        \textbf{Network structure}: Regarding the deep learning model for low-dose CT image noise reduction, the current mainstream methods can be roughly categorized into three types: 

        \smallskip

        \subsubsection{Encoder-decoder}
            The encoder-decoder model uses a symmetrical structural design. 
            Convolutional layers are utilized to form the encoder, carrying out the encoding of spatial information. 
            Then the model uses the same number of deconvolutional layers to form the decoder, which generally fuses feature maps from the encoder using skip connections, such as the REDCNN\cite{Chen_2017} with residual connections, the CPCE\cite{Shan_2018} with conveying-paths connections, etc.
            LDCT images can be denoised through the entire encoder-decoder model.

        \subsubsection{Fully convolution network}
            It means that the whole network is composed of convolution layers. 
            The output images denoised from low-dose CT images are obtained by several layers' convolution operation.
            As for the configuration of the convolution layer, different models have different ideas. 
            Some models just use simple convolutional layers with kernel size set to 5 or 3, such as the denoiser in \cite{Yang_2018}.
            The model in \cite{gholizadeh-ansari_deep_2020} stacks convolutional layers with different dilation rate to increase the receptive field.
            Besides, this type of model also utilizes residual or conveying-paths connections.
            Our proposed EDCNN model is just designed as the FCN structure. 

        \subsubsection{GAN-based algorithms}
            This type of algorithm consists of a generator and a discriminator. 
            The generator is designed as a noise reduction network and can be used alone during the testing stage.
            The discriminator is used to distinguish the denoiser's output and the target high-dose CT image.
            They are optimized in an adversarial strategy with an alternate training process. 
            There are some existing methods\cite{7934380,Yang_2018,choi_real-time_2018,hu_artifact_2019} that use this structure.
            With the further development of GAN, researchers will implement new models and perform experiments to do further exploration.
        
        \smallskip

        In addition to these types, there are also some algorithms that use a multi-model architecture, which uses a cascaded structure\cite{shan_competitive_2019}\cite{ataei2020cascaded} or parallel networks\cite{li2020lowdose}.
        Our paper aims to design a single model and complete the low-dose CT image denoising task efficiently, so this type of algorithms will not be explained in detail here.
        
        \medskip
        
        \textbf{Loss function}: Low dose CT image denoising is one kind of image transformation tasks, and its common loss functions used for optimization are roughly as follows: 

        \smallskip
        \setcounter{subsubsection}{0}

        \subsubsection{Per-pixel loss}
            The goal of low-dose CT image denoising is to get the result close to that using high-dose radiation, so a simple idea is to set the loss function directly as the per-pixel loss between the output image and the target image.
            In this type of loss, the Mean Square Error (MSE) loss function is commonly used\cite{kang_deep_2017}\cite{Chen_2017}. 
            The L1 loss function also belongs to this type.
            However, this kind of loss function has an obvious problem. 
            It can not describe the structure information in the image, which is important in the denoising task. 
            And methods trained by this type tend to output over-smoothed images\cite{Chen_2017}.

        \subsubsection{Perceptual loss}
            In order to solve the spatial information dependence in image transformation tasks, \cite{10.1007/978-3-319-46475-6_43} proposes a new kind of loss function, the perceptual loss.
            It maps images to the feature space and calculates the similarity in this level.
            With regard to mappers, VGGNet\cite{simonyan2014deep} with trained weights is often used.
            With this kind of loss functions, the detail information of the image can be better preserved, but there are also some problems that cannot be ignored, such as the cross-hatch artifacts introduced by this method.

        \subsubsection{Other loss}
            In GAN-based denoising algorithms, the models use adversarial loss during training, inspired by the ideas of DCGAN\cite{radford2015unsupervised}, WGAN\cite{arjovsky2017wasserstein} and so on.
            These loss functions can also capture the structural information of the image to generate more realistic images.
            Except for these types of loss, researchers design some special forms of loss functions. 
            For example, the MAP-NN model in \cite{shan_competitive_2019} proposes the composite loss function which contains three components including adversarial loss, mean-squared error (MSE), and edge incoherence loss.
            Although there are various loss functions, they are all designed to produce images of higher quality.
            
        \smallskip

    \section{Methodology}
        This section presents the proposed edge enhancement-based densely connected network (EDCNN) in detail, including the edge enhancement module, overall model structure, and the loss function used for the optimization process.
        
        \begin{figure*}[tp]
            \centering
            \includegraphics[width=\linewidth]{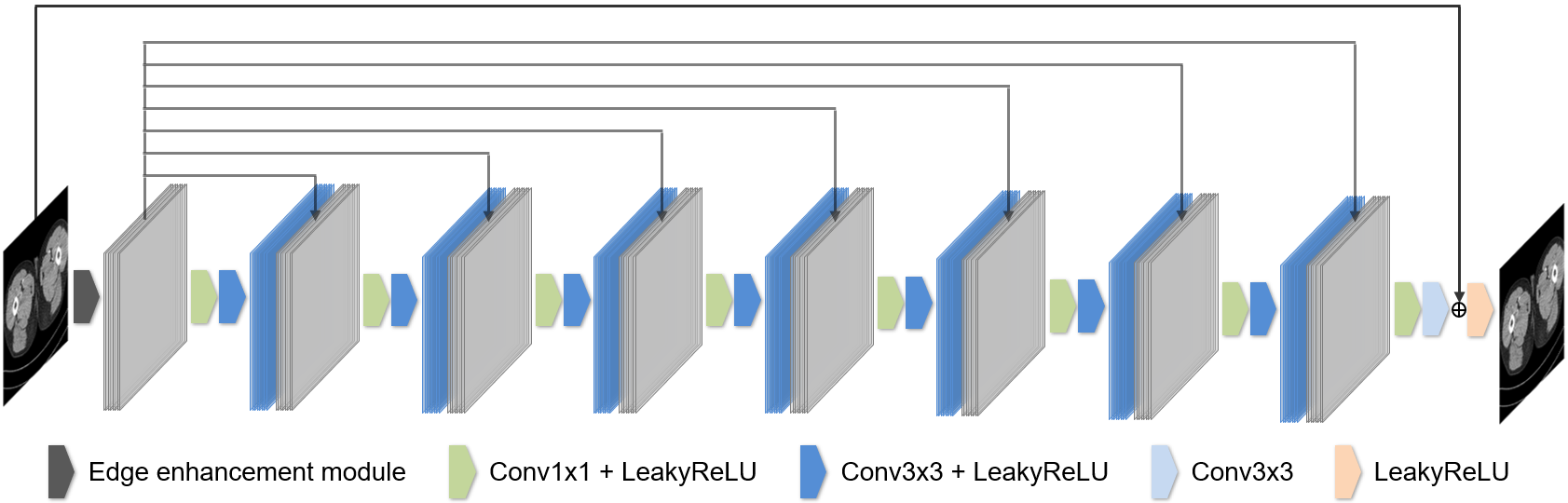}
            \caption{Overall architecture of our proposed EDCNN model.}
            \label{model_structure}
        \end{figure*}

        \subsection{Edge enhancement Module} \label{methodology_section_A}
            Before describing the structure of the whole model, this subsection first introduces the edge enhancement module, which directly acts on the input image.
            
\begin{figure}[h]
                \subfloat[]{\includegraphics[width=0.45\linewidth]{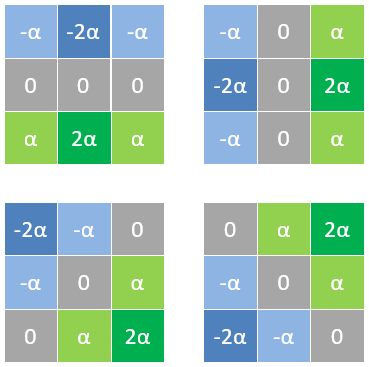} \label{trainable_sobel}}
                \hspace{0.05\linewidth}
                \subfloat[]{\includegraphics[width=0.45\linewidth]{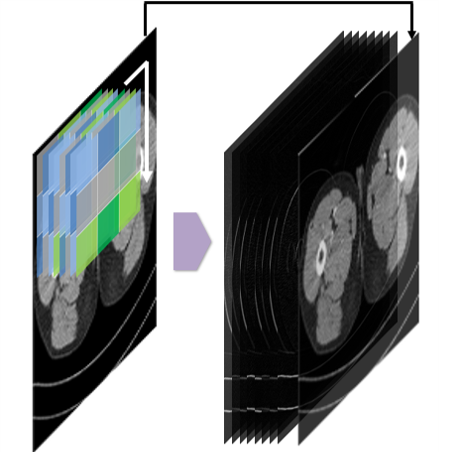} \label{edge_enhancement_module}}
                \caption{The designed edge enhancement module. (a) Four types of trainable Sobel operators; (b) Process of this module.}
                \label{trainable_sobel_and_edge_enhancement_module}
            \end{figure}

            In this module, we design the trainable Sobel convolution. 
            As shown in Fig.~\ref{trainable_sobel}, different from the traditional fixed-value Sobel operator\cite{sobel_operator}, a learnable parameter $\alpha$ is defined in the trainable Sobel operator, which is called Sobel factor by us.
            The value of this parameter can be adaptively adjusted during the optimization of training process, so it can extract edge information of different intensity.
            Besides, we define four types of operators as a group (Fig.~\ref{trainable_sobel}), including vertical, horizontal, and diagonal directions. 
            Multiple groups of trainable Sobel operators can be used in this module. 

            In the flow of this module (Fig.~\ref{edge_enhancement_module}), firstly, it uses a certain number (a multiple of 4) of trainable Sobel operators on the input CT image, performing convolution operations to obtain a set of feature maps for extracting edge information.
            And then the module stacks them with the input low-dose CT Images together in the channel dimension to get the final output of this module. 
            The goal of this module is to enrich the input information of the model at the level of data source and strengthen the effect of edge information to the model.

        \subsection{Overall Network Architecture} \label{methodology_section_B}
            The proposed network architecture is illustrated in Fig.~\ref{model_structure}, which is called Edge enhancement based Densely connected Convolutional Neural Network (EDCNN).
            The whole model consists of an edge enhancement module and eight convolution blocks. 
            The edge enhancement module has been explained in Section~\ref{methodology_section_A}. 
            The number of trainable Sobel operators we use is 32 (8 groups of the four types).
            
            As for the model structure after the edge enhancement module, the purpose of our design is to retain the image details in the process as much as possible. 
            Inspired by the DenseNet\cite{huang2016densely}, we design a low-dose CT denoising model with dense connection, trying to make full use of the extracted edge information and the original input.
            Specifically, as shown by the line in Fig.~\ref{model_structure}, we convey the output of the edge enhancement module to each convolution block through skip connection and concatenate them in the channel dimension.
            The inner structure of the latter convolution blocks is exactly the same except for the last layer.
            These blocks are composed of 1x1 and 3x3 convolution, and the number of convolutional filters is all set to 32.
            The number of 3x3 convolutional filters in the last layer is 1, corresponding to the output of a single channel. 
            In each block, the point-wise convolution with 1x1 kernels is used to fuse the outputs of the previous layer and edge enhancement module, and the convolution with 3x3 kernels is used to learn features in the image as usual.
            Besides, to keep the output size and input size the same, feature maps in the model are padded to ensure that the spatial size does not change during the forward propagation.

            In order to accelerate the convergence of the model and simplify the task of the main structure of model, we let the model directly learn the noise distribution and reconstruction information. 
            So the output of the last convolution block is added with the original low-dose CT image to get the final noise-denoised images.
            In Fig.~\ref{model_structure}, the top line represents this residual connection, and the symbol, which consists of a circle and a plus sign, represents the element-wise addition.

            \subsection{Compound Loss Function} \label{compound_loss_function}
                The ultimate goal of CT image denoising is to obtain the output results similar to target images with a higher dose of radiation exposure.
                Assuming that $I_{LDCT} \in \mathbb{R}^{1 \times w \times h}$ represents an LDCT image with the size of $w \times h$, and $I_{NDCT} \in \mathbb{R}^{1 \times w \times h}$ represents the target NDCT image, the denoising task can be expressed as follows:
                \begin{equation}
                    F(I_{LDCT}) = I_{Output} \approx I_{NDCT} \label{eq_task}
                \end{equation}
                where $F$ represents the noise reduction method, and $I_{Output}$ denotes the output image of the denoiser.

                To achieve this purpose, the MSE (Eq.~\ref{eq_mse_loss}) is widely used in previous methods as the loss function.
                The distance between the model's output and the target image is calculated pixel by pixel. 
                However, the loss has been verified by lots of experiments, tending to make output images over-smoothed and increase the image blur.

                In order to overcome the problem, this paper introduces the compound loss function, which fuses MSE loss and multi-scales perceptual loss, as shown in the following formulas: 
                \begin{equation}
                    L_{mse} = \frac{1}{N} \sum_{i=1}^N \left\| F\left(x_{i}, \theta\right) - y_{i} \right\|^{2} \qquad\qquad\qquad\qquad\quad \label{eq_mse_loss}
                \end{equation}
                \begin{small}
                    \begin{equation}
                        L_{multi-p} = \frac{1}{NS} \sum_{i=1}^{N} \sum_{s=1}^S \left\|\phi_s\left(F\left(x_{i}, \theta\right), \hat{\theta}\right)-\phi_s\left(y_{i}, \hat{\theta}\right)\right\|^{2} \label{eq_perceptual_loss}
                    \end{equation}
                \end{small}
                \begin{equation}
                    L_{compound} = L_{mse} + w_{p} \cdot L_{multi-p} \qquad\qquad\qquad\quad ~ \label{eq_compound_loss}
                \end{equation}
                In these formulas, we use $x_{i}$ as the input, $y_{i}$ as the target, and $N$ is the number of images.
                Same as above, $F$ represents the noise reduction model with parameters $\theta$.
                In Eq.~\ref{eq_perceptual_loss}, the symbol $\phi$ represents the model with fixed pre-trained weights $\hat{\theta}$, which is used to calculate the perceptual loss.
                And $S$ is the number of scales. 
                The $w_{p}$ in Eq.~\ref{eq_compound_loss} denotes the weight of the second part of the compound loss function.

\begin{figure}[t]
                    \includegraphics[width=\linewidth]{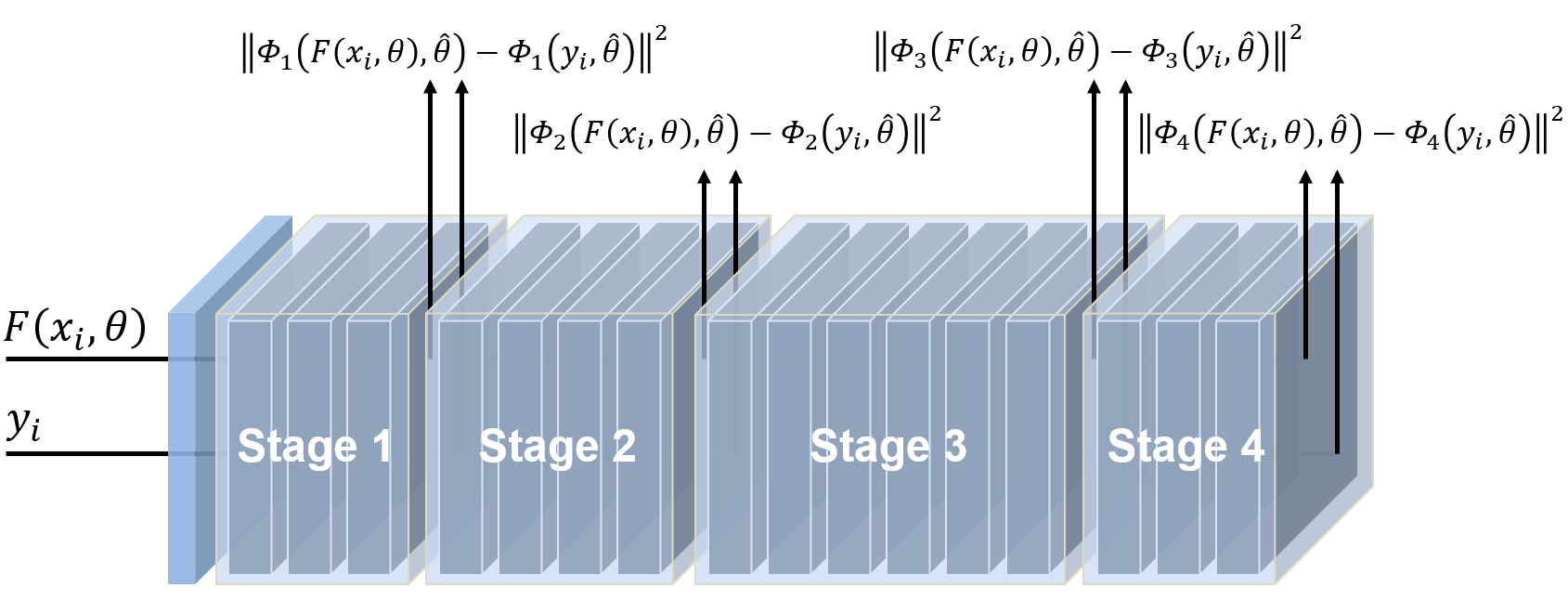}
                    \caption{The multi-scales perceptual loss. It is based on ResNet-50, which contains 4 main stages. For more details about this model, please refer to paper \cite{He_2016_CVPR}}
                    \label{multi_scale_perceptual_loss}
                \end{figure}

                Regarding the perceptual loss, as shown in Fig.~\ref{multi_scale_perceptual_loss}, we utilize the ResNet-50\cite{He_2016_CVPR} as the feature extractor to get the multi-scale perceptual loss.
                Specifically, we discard the pooling layer and the fully connected layer at the end of the model, retaining only the convolution layers in the front of this model.
                In the beginning, we first load the model's weights trained on the ImageNet dataset\cite{5206848}, and then freeze these weights during training.
                When calculating perceptual loss value, both the denoised output and target image are sent to the extractor to do forward propagation (Fig.~\ref{multi_scale_perceptual_loss}).
                We choose the feature maps after four stages of ResNet, in each of that the spatial scale of the image will be halved, representing feature spaces of different scales.
                Then we use the MSE to measure the similarity of these feature maps.
                The multi-scales perceptual loss is obtained by averaging these values.

                By combining MSE and multi-scales perceptual loss, we can concern both the per-pixel similarity and the structural information of CT images.
                And we can adjust the hyperparameter $w_{p}$ to balance the two loss components (Eq.~\ref{eq_compound_loss}).

    \section{Experiments and Results}
        This section explains the dataset used to train and test the proposed model, the configuration of the experiment. 
        And then we show the experimental results in this section, evaluating the noise reduction performance of the model.

        \subsection{Dataset}
            In the experiment of our study, we utilize the dataset of the 2016 NIH AAPM-Mayo Clinic Low-Dose CT Grand Challenge\cite{mccollough_low-dose_2017}, which is used by current mainstream methods in the field of low-dose CT image denoising.
            It contains the paired normal-dose CT (NDCT) images and synthetic quarter-dose CT images (LDCT) with a size of 512x512 pixels, collected from 10 patients.
            So there are LDCT images for inputs of the model and NDCT images as targets, which can support the supervised training process.

            As for the data preparation, we split the dataset before training, using nine patients' CT images as the training set, and the rest one patient's images as the test set.
            
        \subsection{Experimental Setup}
            The structure of the model and number of filters in each layer have been described in Section~\ref{methodology_section_B}, which is implemented by us based on the Pytorch framework\cite{NIPS2019_9015}. 
            We use the default random initialization for the convolution layers in this model, and the Sobel factors of all edge enhancement modules are initialized to 1 before training.
            Besides, the hyperparameter $w_{p}$ of the compound loss function is set to 0.01.

            During training, we apply a data augmentation strategy that crops patch randomly.
            Specifically, 4 patches with a size of 64x64 pixels will be randomly cropped from one LDCT image, and the input batch we used is taken from 32 images, which has 128 patches in total, so is the target batch of NDCT images.
            In the process of optimization, we utilize the AdamW optimizer\cite{loshchilov2018fixing} with the default configuration.
            We set the learning rate to 0.001, and conduct 200 epochs of training to make the model converge.
            When testing the model, because of the model's fully convolutional structure, there is no limit on the size of the input image. 
            So we let the trained model use LDCT images with the size of 512x512 pixels as the input and directly outputs the denoised results.

        \subsection{Results}
            \begin{figure*}[t]
                \subfloat[LDCT]{\includegraphics[width=0.33\linewidth]{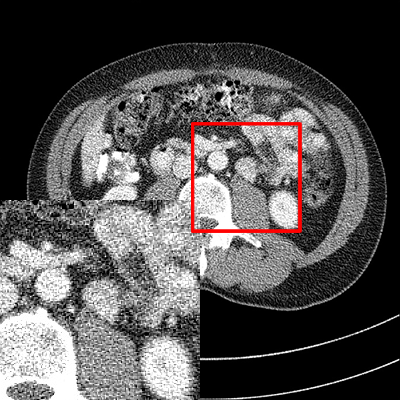} \label{test_img_50_ldct}}
                \hspace{0.01\linewidth}
                \subfloat[NDCT]{\includegraphics[width=0.33\linewidth]{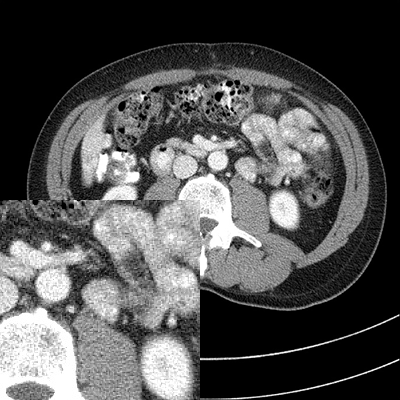} \label{test_img_50_ndct}}
                \hspace{0.01\linewidth}
                \subfloat[REDCNN]{\includegraphics[width=0.33\linewidth]{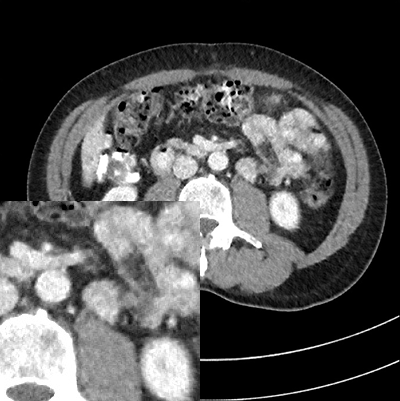} \label{test_img_50_pred_REDCNN}}
                \\
                \subfloat[WGAN]{\includegraphics[width=0.33\linewidth]{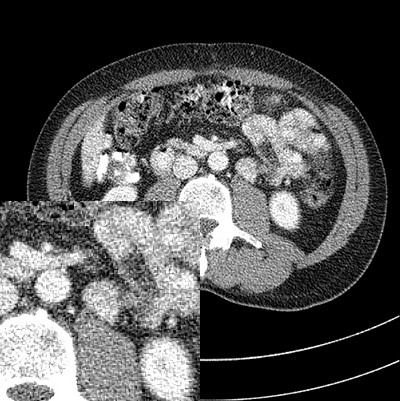} \label{test_img_50_pred_WGAN}}
                \hspace{0.01\linewidth}
                \subfloat[CPCE]{\includegraphics[width=0.33\linewidth]{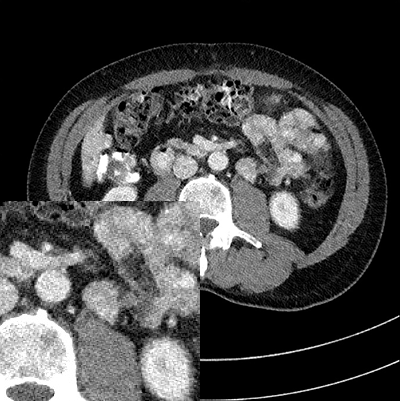} \label{test_img_50_CPCE}}
                \hspace{0.01\linewidth}
                \subfloat[EDCNN]{\includegraphics[width=0.33\linewidth]{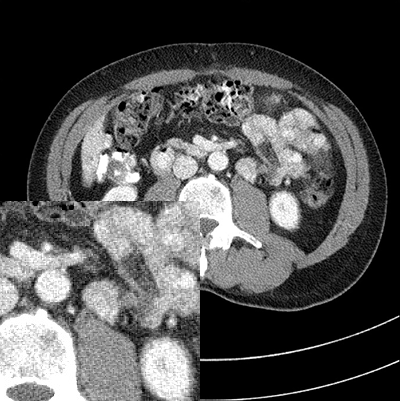} \label{test_img_50_pred_EDCNN}}
                \caption{The denoised results of different model. The Region of Interest (RoI) in the red box is selected and magnified in the lower left corner of images for a clearer comparison.}
                \label{denoised_result_comparison}
            \end{figure*}

            \begin{table*}[t]
                \centering
                \fontsize{8}{12}\selectfont
                \begin{threeparttable}
                    \caption{Quantitative Comparison among Different Models on the AAPM Dataset}
                    \label{comparison_of_models}
                    \begin{tabular}{|l|c|c|c|c|c|c|c|c|}
                        \hline
                        \multirow{2}{*}{\textbf{~~~Method}} & \multicolumn{4}{|c|}{\textbf{Loss}} & \multicolumn{4}{|c|}{\textbf{Metric}} \cr
                        \cline{2-9} 
                        & \textbf{\textit{MSE Loss}} & \textbf{\textit{VGG-P Loss}} & \textbf{\textit{Adversarial Loss}} & \textbf{\textit{MS-P Loss}} & \textbf{\textit{PSNR}}& \textbf{\textit{SSIM}}& \textbf{\textit{RMSE}}& \textbf{\textit{VGG-P}} \cr
                        \hline
                        \hline
                        LDCT   & - & - & - & - & 36.7594$\pm$0.9675 & 0.9465$\pm$0.0113 & 0.0146$\pm$0.0016 & 0.0377$\pm$0.0055 \cr
                        \hline
                        REDCNN\cite{Chen_2017} & \checkmark & - & - & - & \textcolor{red}{\bf 42.3891$\pm$0.7613} & \textcolor{blue}{\bf 0.9856$\pm$0.0029} & \textcolor{red}{\bf 0.0076$\pm$0.0007} & 0.0218$\pm$0.0048 \cr
                        \hline
                        WGAN\cite{Yang_2018} & - & \checkmark & \checkmark & - & 38.6043$\pm$0.9492 & 0.9647$\pm$0.0078 & 0.0108$\pm$0.0013 & 0.0072$\pm$0.0019 \cr
                        \hline
                        CPCE\cite{Shan_2018} & - & \checkmark & \checkmark & - & 40.8209$\pm$0.7905 & 0.9740$\pm$0.0050 & 0.0093$\pm$0.0009 & \textcolor{red}{\bf 0.0043$\pm$0.0011} \cr
                        \hline
                        EDCNN  & \checkmark & - & - & \checkmark & \textcolor{blue}{\bf 42.0835$\pm$0.8100} & \textcolor{red}{\bf 0.9866$\pm$0.0031} & \textcolor{blue}{\bf 0.0079$\pm$0.0007} & \textcolor{blue}{\bf 0.0061$\pm$0.0014} \cr
                        \hline
                    \end{tabular}
                    \begin{tablenotes}
                        \item []~~The `VGG-P' means perceptual loss based on VGGNet, and `MS-P Loss' represents the multi-scales perceptual. PSNR, SSIM, RMSE and VGG-P are used as the metric, which are shown in the form of $mean \pm std$. The best ones are marked in red, and the second best ones are marked in blue.
                    \end{tablenotes}
                \end{threeparttable}
            \end{table*}

            This subsection shows the noise reduction results of our model.
            For fairness, we choose the REDCNN\cite{Chen_2017}, WGAN\cite{Yang_2018} and CPCE\cite{Shan_2018} for comparison, because of their design of the single model, which is the same as our proposed model. 
            These models also adopt the structure of convolutional neural networks, but each of them has its characteristics. 
            We re-implement these models, training them on the same training set.
            The left part of Table.~\ref{comparison_of_models} shows the configuration of loss functions used by them, including our model as well.

            In the noise reduction task, there are three common criteria for quantitative analysis a model, including the Peak Signal to Noise Ratio (PSNR), Structural SIMilarity (SSIM), and Root Mean Square Error (RMSE).
            Besides, we add a metric, VGG-P, which is the commonly used perceptual loss based on VGGNet19\cite{simonyan2014deep}, measuring the distance in the final convolution layer's feature space\cite{10.1007/978-3-319-46475-6_43}. 
            As shown on the right part of Table.~\ref{comparison_of_models}, all the models are tested on the split test set of the AAPM Challenge's dataset.
            We calculate and count the mean and standard deviation of these metrics.
            Through this table, we can find that the REDCNN based on MSE loss has the best performance on the metrics of PSNR and RMSE.
            By using perceptual loss based on VGGNet, the WGAN and CPCE have a good result on VGG-P.
            As for our proposed EDCNN, based on compound loss, it achieves the best or suboptimal results on every criterion, which can balance the per-pixel and structure-wise performance.

            Since the calculation process of PSNR and RMSE is directly related to MSE, the model trained by just using MSE as the loss function can get good results on these metrics.
            However, these criteria can not truly reflect the visual quality of the output image, so they can only be used as a relative reference.
            For comparison of denoised results, as illustrated in Fig.~\ref{denoised_result_comparison}, we choose a CT image with a complex structure to show the performance of these models.
            We can notice that there is more noise in LDCT images (Fig.~\ref{test_img_50_ldct}) than in NDCT images (Fig.~\ref{test_img_50_ndct}).
            After denoising from the LDCT image, the REDCNN's output (Fig.~\ref{test_img_50_pred_REDCNN}) is obviously over-smoothed.
            Although it has the highest PSNR and the lowest RMSE, the visual perception of the image is not good, which has the problem of image blur and loss of structure details.
            The WGAN and CPCE are all based on Wasserstein GAN with perceptual loss and adversarial loss.
            Fig.~\ref{test_img_50_pred_WGAN} shows the denoised CT image of WGAN, which retains the structural information of the original image, but its suppression of noise is still relatively poor.
            Shown in Fig.~\ref{test_img_50_CPCE} and Fig.~\ref{test_img_50_pred_EDCNN}, the CPCE model and our EDCNN have comparable performance.
            The output images of them are all much similar to the target NDCT image (Fig.~\ref{test_img_50_ndct}), preserving the subtle structure of the CT image.
            But from the details of the noise dots, we can still notice the difference between them.
            The EDCNN has better noise reduction performance than CPCE, which is also consistent with the value of the metrics in Table.~\ref{comparison_of_models}.

            \begin{table}[h]
                \centering
                \fontsize{7.75}{11}\selectfont
                \begin{threeparttable}
                    \caption{Subjective Scores on Image Quality}
                    \label{score_of_models}
                    \begin{tabular}{|l|c|c|c|}
                        \hline
                        \multirow{2}{*}{\textbf{~Method}} & \multicolumn{3}{|c|}{\textbf{Score}} \cr
                        \cline{2-4} & \textbf{\textit{Noise Reduction}}& \textbf{\textit{Structure Preservation}}& \textbf{\textit{Overall Quality}} \cr
                        \hline
                        \hline
                        REDCNN & \textcolor{red}{\bf 4.19$\pm$0.19} & 3.06$\pm$0.22 & 3.15$\pm$0.44 \cr
                        \hline
                        WGAN   & 2.36$\pm$0.34 & 3.58$\pm$0.17 & 3.49$\pm$0.16 \cr
                        \hline
                        CPCE   & 3.45$\pm$0.17 & \textcolor{blue}{\bf 4.03$\pm$0.19} & \textcolor{blue}{\bf 3.95$\pm$0.42} \cr
                        \hline
                        EDCNN  & \textcolor{blue}{\bf 3.64$\pm$0.12} & \textcolor{red}{\bf 4.07$\pm$0.21} & \textcolor{red}{\bf 4.13$\pm$0.20} \cr
                        \hline
                    \end{tabular}
                    \begin{tablenotes}
                        \item []~~The scores are shown in the form of $mean \pm std$ (Perfect score is 5 points). The best ones are marked in red, and the second best ones are marked in blue.
                    \end{tablenotes}
                \end{threeparttable}
            \end{table}

            In order to obtain the quantitative visual evaluation, we conduct the blind reader study.
            Specifically, we select 20 groups of models' denoised results in the test set with different body parts.
            Each group includes six CT images.
            The LDCT and NDCT images are used as references, and the other four images are the outputs of the above four models, which are randomly shuffled in each group.
            Readers are asked to score the denoised CT images on three levels, including noise reduction, structure preservation and the overall quality, with a full score of 5 points for each item.
            As shown in Table.~\ref{score_of_models}, we present the statistics of the subjective scores in the form of $mean \pm std$.
            The REDCNN has the best performance of noise reduction, and the GAN-based WGAN and CPCE have a high score in structure preservation.
            Concerning our designed EDCNN model, it consider both noise reduction and structure preservation because of the compound loss. In addition, the EDCNN gets a high score in overall image quality.
            
        \subsection{Ablation Study}
            In this part, we compare and analyze the performance of our model under different configurations of model structure and loss function.
            And we discuss the validity of the final design in our proposed EDCNN model.

            \medskip

            \subsubsection{Structure and Module}
                To explore the effect of each component of EDCNN model, we make a decomposition experiment on the structure. 
                First, we designed a basic model (BCNN), removing the dense connection and edge enhancement module from the structure shown in Fig.~\ref{model_structure}, and then we add dense connection (BCNN+DC) and edge enhancement module (BCNN+DC+EM, EDCNN) in turn.
                In order to fully demonstrate the potential capacity of the model, all models are trained with MSE loss with the same training strategy.

                \begin{figure}[h]
                    \includegraphics[width=\linewidth]{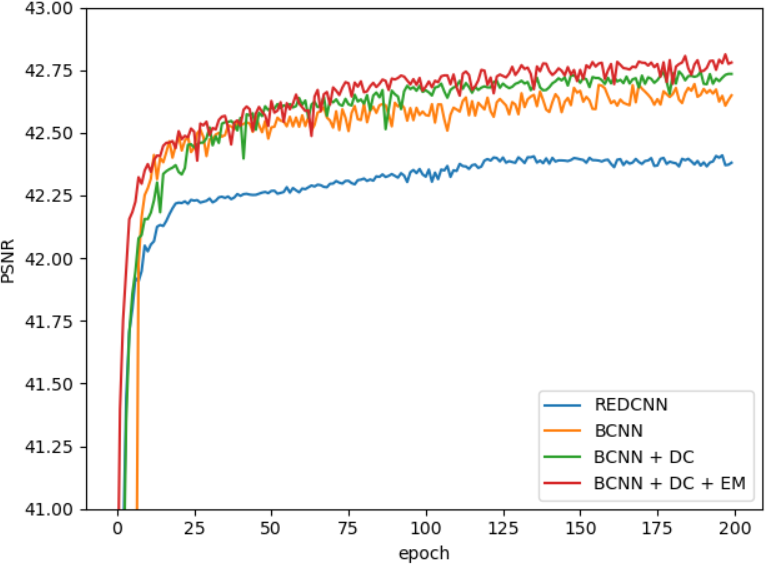}
                    \caption{The PSNR curves in the training process.}
                    \label{psnr_curve}
                \end{figure}

                \begin{table}[h]
                    \centering
                    \fontsize{7.75}{10}\selectfont
                    \begin{threeparttable}
                        \caption{Performance Comparison on Model Structure}
                        \label{structure_comparison}
                        \begin{tabular}{|l|c|c|c|}
                            \hline
                            \multirow{2}{*}{\textbf{~~~~~~~~Method}} & \multicolumn{3}{|c|}{\textbf{Metric}} \cr
                            \cline{2-4} & \textbf{\textit{PSNR}}& \textbf{\textit{SSIM}}& \textbf{\textit{RMSE}} \cr
                            \hline
                            \hline
                            LDCT & 36.7594$\pm$0.9675 & 0.9465$\pm$0.0113 & 0.0146$\pm$0.0016 \cr
                            \hline
                            REDCNN & 42.3891$\pm$0.7613 & 0.9856$\pm$0.0029 & 0.0076$\pm$0.0007 \cr
                            \hline
                            BCNN & 42.6654$\pm$0.7929 & 0.9864$\pm$0.0027 & 0.0074$\pm$0.0007 \cr
                            \hline
                            BCNN+DC & 42.7444$\pm$0.7684 & 0.9868$\pm$0.0025 & {\bf 0.0073$\pm$0.0007} \cr
                            \hline
                            BCNN+DC+EM & {\bf 42.8128$\pm$0.7726} & {\bf 0.9870$\pm$0.0025} & {\bf 0.0073$\pm$0.0007} \cr
                            \hline
                        \end{tabular}
                        \begin{tablenotes}
                            \item []~DC represents Dense Connection, and EM represents the Edge enhancement Module. The best ones on each metric are marked in bold ($mean \pm std$).
                        \end{tablenotes}
                    \end{threeparttable}
                \end{table}

                Fig.~\ref{psnr_curve} shows the curves of PSNR, testing on the test set for trained models at each epoch.
                We also add REDCNN as a comparison.
                It is worth noting that the basic model (BCNN) of our design already achieves better performance than REDCNN.
                And the value of PSNR will increase continuously by adding the dense connection and edge enhancement module.
                Besides, the edge enhancement module accelerates the convergence process of the model.
                In Table.~\ref{structure_comparison}, we can check the value of PSNR, SSIM, RMSE for these models.
                And the complete EDCNN model has the best results on these metrics.

            \medskip
            
            \subsubsection{Models of Perceptual Loss}
                As demonstrated in Section~\ref{compound_loss_function}, the model chosen to calculate the perceptual loss in our method is the ResNet-50.
                Regarding the model of perceptual loss, we compare the ResNet-50 with the VGGNet-19 that is commonly used by existing methods.
                In this experiment, we just train the EDCNN model by single perceptual loss.
                According to the previous methods, we use the last convolution layer's ouput of VGGNet-19 to calculate the loss.
                As for the ResNet-50 we used, we also utilize the feature maps of its last convolution layer with the same idea for comparison.
                
                \begin{figure}[h]
                    \subfloat[LDCT]{\includegraphics[width=0.24\linewidth]{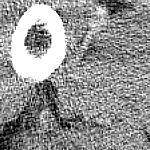} \label{Crop_test_img_1_ldct}}
                    \subfloat[VGG-P]{\includegraphics[width=0.24\linewidth]{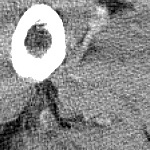} \label{Crop_test_img_1_vgg}}
                    \subfloat[ResNet-P]{\includegraphics[width=0.24\linewidth]{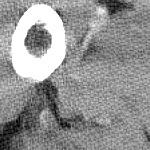} \label{Crop_test_img_1_resnet}}
                    \subfloat[NDCT]{\includegraphics[width=0.24\linewidth]{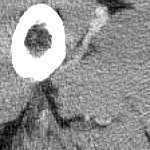} \label{Crop_test_img_1_ndct}}
                    \caption{The denoised results of EDCNN with different perceptual loss. VGG-P and ResNet-P represent the perceptual loss based on VGGNet and ResNet respectively.}
                    \label{perceptual_model_comparison}
                \end{figure}

                Models optimized by perceptual loss tend to ouput images with some kind of texture-like noise.
                By observing Fig.~\ref{perceptual_model_comparison} carefully, we can find that the noise graininess of Fig.~\ref{Crop_test_img_1_vgg} is bigger than that of Fig.~\ref{Crop_test_img_1_resnet}. 
                And from the visual appearance, Fig.~\ref{Crop_test_img_1_resnet} is closer to the NDCT image (Fig.~\ref{Crop_test_img_1_ndct}).
                So we use the ResNet-50 model in our perceptual loss function, which has stronger feature extraction ability than the VGGNet.

            \medskip
            
            \subsubsection{Multi-Scales Perceptual Loss}
\begin{figure}[t]
                    \subfloat[LDCT]{\includegraphics[width=0.155\linewidth]{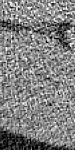} \label{Crop_test_img_100_ldct}}
                    \subfloat[S-4]{\includegraphics[width=0.155\linewidth]{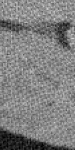} \label{Crop_test_img_100_s4}}
                    \subfloat[S-43]{\includegraphics[width=0.155\linewidth]{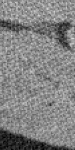} \label{Crop_test_img_100_s43}}
                    \subfloat[S-432]{\includegraphics[width=0.155\linewidth]{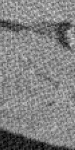} \label{Crop_test_img_100_s432}}
                    \subfloat[S-4321]{\includegraphics[width=0.155\linewidth]{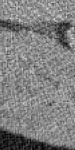} \label{Crop_test_img_100_s4321}}
                    \subfloat[NDCT]{\includegraphics[width=0.155\linewidth]{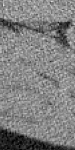} \label{Crop_test_img_100_ndct}}
                    \\
                    \subfloat[LDCT]{\includegraphics[width=0.19\linewidth]{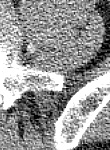} \label{Crop_test_img_15_ldct}}
                    \subfloat[MSE]{\includegraphics[width=0.19\linewidth]{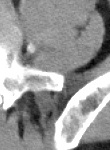} \label{Crop_test_img_15_mse}}
                    \subfloat[MS-P]{\includegraphics[width=0.19\linewidth]{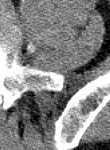} \label{Crop_test_img_15_resnet}}
                    \subfloat[Compound]{\includegraphics[width=0.19\linewidth]{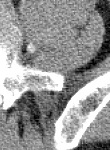} \label{Crop_test_img_15_compound}}
                    \subfloat[NDCT]{\includegraphics[width=0.19\linewidth]{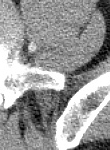} \label{Crop_test_img_15_ndct}}
                    \caption{Comparison of denoised images with different configuration of loss. (a)-(f) show different setup of multi-scales loss, and (g)-(k) compare the single loss with our compound loss.}
                    \label{loss_comparison}
                \end{figure}

                When using perceptual loss, we need to decide which layer of feature maps to use.
                Here we explore the different combinations of the multi-scales perceptual loss.
                Specifically, we utilize the output features of four stages in ResNet-50 (Fig.~\ref{multi_scale_perceptual_loss}).
                Four types of loss functions are designed, including perceptual loss with S-4, S-43, S-432, and S-4321.
                `S' represents the stage and the numbers denote the number of stages used to get feature maps.
                The loss function will calculate MSE on these stages' extracted features (Eq.~\ref{eq_perceptual_loss}), and compute their average to get the final loss value. 
                Fig.~\ref{loss_comparison}(a-f) show the output images, we can find that as the number of stages used increases, the `texture' of the denoised result is closer to the NDCT image.
                Therefore, we decide to use the output features of four stages in ResNet-50 model to calculate our method's perceptual loss.

            \medskip

            \subsubsection{Single or Compound Loss}
                During this experiment, we attain three EDCNN models trained on single MSE loss, single multi-scales perceptual loss, and compound loss respectively.
                They are trained in the same way, except for their loss.

                As shown in Fig.~\ref{loss_comparison}(g-k), we can compare the visual quality among these denoised CT images.
                Apparently, the result of MSE-based EDCNN model (Fig.~\ref{Crop_test_img_15_mse}) is already over-smoothed, missing too much detail for later diagnosis, and it brings difficulties for doctors to make judgments.
                As for Fig.~\ref{Crop_test_img_15_resnet} and Fig.~\ref{Crop_test_img_15_compound}, they show similar quality in detail retention, which further verifies the effectiveness of the multi-scales perceptual loss.
                In the meanwhile, we can notice that Fig.~\ref{Crop_test_img_15_compound} is slightly clearer than Fig.~\ref{Crop_test_img_15_resnet}.
                The latter introduces some visible artifacts.
                EDCNN based on compound loss has better performance.

    \section{Conclusion}
        In summary, this article presents a new denoising model with a densely connected convolutional architecture, the Edge enhancement-based Densely Connected Network (EDCNN). 
        Through the designed edge-enhancement module based on trainable Sobel operators, the method can get richer edge information of the input image adaptively.
        Besides, we introduce the compound loss function, which is a weighted fusion of MSE loss and multi-scales perceptual loss.
        Using the well-known mayo dataset, we make a lot of experiments and our method achieves better performance compared with previous models.
        In the future, we plan to further explore the multi-models structure based on our proposed EDCNN model and extend it to other image transformation tasks.

    \section*{Acknowledgment}
        This work is supported by the Institute of Information Science in Beijing Jiaotong University, 
        and we gratefully acknowledge the support from its laboratory for providing us the Titan GPUs to accomplish this research.

    \bibliography{references}

\end{document}